# X-ray tomography investigation of cyclically sheared granular materials

Yi Xing[1], Jie Zheng[1], Jindong Li[1], Yixin Cao[1], Wei Pan[1], Jie Zhang[1,2] and Yujie Wang[1,3*]

[1]*School of Physics and Astronomy, Shanghai Jiao Tong University, 800 Dong Chuan Road, Shanghai 200240, China*

[2]*Institute of Natural Sciences, Shanghai Jiao Tong University, Shanghai 200240, China*

[3]*Materials Genome Initiative Center, Shanghai Jiao Tong University, 800 Dong Chuan Road, Shanghai 200240, China*

We perform combined X-ray tomography and shear force measurements on a cyclically sheared granular system with highly transient behaviors, and obtain the evolution of microscopic structures and the macroscopic shear force during the shear cycle. We explain the macroscopic behaviors of the system based on microscopic processes, including the particle level structural rearrangement and frictional contact variation. Specifically, we show how contact friction can induce large structural fluctuations and cause significant shear dilatancy effect for granular materials, and we also construct an empirical constitutive relationship for the macroscopic shear force.

Dense granular materials exhibit complex behaviors upon shear. Notable examples include significant dilatancy [1], formation of shear bands [2], emergence of anisotropic force networks [3] and critical state [4]. However, the theoretical understanding of sheared granular materials remains very challenging since conventional continuum mechanics are no longer applicable due to the disordered nature of granular materials, as well as the complex role friction plays [5,6]. In practice, empirical constitutive theories are normally employed, including the famous

Mohr-Coulomb criterion [7] and more sophisticated visco-plastic models [8]. However, these models are mainly macroscopic ones with little microscopic basis despite the fact that it is well-known experimentally that microscopic structure and dynamics can strongly influence the macroscopic response [9]. Granular materials belong in general to the family of disordered materials like metallic and colloidal glasses, foams, emulsions, *etc*., and physicists have tried to treat the flow of dense granular materials within the general framework of disordered materials [9]. For example, theories based on free volume [10], shear transformation zone (STZ) [11] and soft glass rheology (SGR) theories [12], which are built upon microscopic and mesoscopic information to predict the plastic behaviors of certain disordered materials, have been directly applied to the granular case [13,14]. However, granular materials possess some peculiar properties, *e.g.*, force chains [15,16], significant volume and stress fluctuations [6,17,18], and complex microscopic dynamics [19], which are not shared among all disordered materials. These properties are mainly induced by the frictional inter-particle contacts. Therefore, to construct the correct constitutive theory for the flow of dense granular materials, physics on particle and contact levels have to be included [20].

In this letter, we apply cyclic shear to a three-dimensional (3D) disordered granular system. The system displays highly transient dilatational behavior and mechanical response since the system always has to reorganize upon sudden shear reversal [14,21]. By combing X-ray tomography and shear force measurements, we obtain both the microscopic structure evolution and macroscopic mechanical response of the system simultaneously. Based on these results, we can explain the macroscopic behaviors of sheared granular matter through the microscopic structural information including both the particle level structural rearrangement and frictional contact evolution. We can now understand why granular materials demonstrate significant shear dilation as compared to other disordered materials, and we can also construct an empirical constitutive relationship to account for the macroscopic shear force behavior.

Our granular system consists of bi-disperse smooth plastic beads ($\mu = 0.4$, diameters 5 and 6 mm, 7000 particles of each size). The bi-disperse nature ensures no crystallization occurs in the system. The particles are placed in a shear cell mounted on a linear stage as shown in Fig. 1(a). The shear cell is in simple shear geometry and has an inner dimension of $24d \times 24d \times 24d$, where $d$=5mm is the small particle's diameter and is set as the unit length. A lid that can freely translate vertically is placed on top of the particle packing (3.6kg, about 3 times of the total particle weight) to maintain a constant pressure and reduce the vertical pressure gradient. Shear is generated by a stepping motor attached to the bottom plate of the shear cell through a linear stage. A force sensor (IMADA ZTS-50N) connecting the motor and the stage is used to measure the shear force.

We prepare the initial state of the system by applying thousands of cyclic shear cycles which ensures that a steady state is reached [19]. The shear strain amplitude we apply is $\gamma = 0.167$ and the strain rate is $\dot{\gamma} = 0.25 \text{s}^{-1}$. The corresponding inertial number is $I = \dot{\gamma}d / \sqrt{P/\rho} = 3 \times 10^{-4}$, suggesting that the system is in a quasi-static regime. After the initial preparation, we divide one shear cycle into 80 steps and an X-ray tomography scan is performed after each shear step through a medical CT scanner (UEG Medical Imaging Equip. Co. Ltd., 0.2 mm spatial resolution). Following similar image processing procedures as previous studies [22], the sizes and positions of all particles are obtained with an error less than $3\times10^{-3}d$ through a maker-based watershed segmentation technique. Based on an in-house tracking algorithm [23], we can then extract the trajectories and contact geometries of all particles. Simultaneously, the shear force is measured at a frequency of $0.23s^{-1}$ and with an accuracy of 0.1N. For better statics, the measurements are repeated for consecutive 12 shear cycles. More details about the experimental setup and image processing procedures can be found in [23].

We calculate the global volume fraction $\phi$ based on radical Voronoi tessellation [24]: $\phi = \sum V_g / \sum V_{voro}$, where $V_g$ and $V_{voro}$ are each particle's volume and its corresponding Voronoi cell volume. To exclude the boundary effect, our structural analysis only involves particles that are at least $2d$ away from the boundary of the shear cell. Fig. 1(b) shows the evolution of $\phi$ and shear force $F$ of the system within one cycle. Starting from the initial symmetric state (shear strain $\gamma=0$), $\phi$ decreases as the strain $\gamma$ increases, then $\phi$ returns to almost the same initial value as the shear is totally reversed with small hysteresis. The similar behavior is repeated for the other half shear cycle. While for the shear force $F$, starting from the symmetric state, it increases as the strain $\gamma$ increases; at the shear reversal point, $F$ changes its direction and its magnitude drops abruptly to almost zero, then it increases monotonically till the next shear reversal point.

Empirical constitutive relations are ordinarily employed to account for macroscopic behaviors, *e.g.*, $\phi$ and $F$, of sheared granular materials [8]. However, it is important to develop theories which construct these relations based on microscopic information [25]. Following this approach, we first characterize the microscopic structure of our system by Delaunay network and then extract the microscopic origin of the macroscopic behaviors by analyzing the structural and topological evolution of the Delaunay network upon shear.

Delaunay tessellation partitions the packing structure into non-overlapping tetrahedra, whose vertices are the centers of four neighboring particles. Upon shear, as the structure of the system varies, the corresponding Delaunay tetrahedra distort gradually and get destroyed when local neighbor switching process occurs, which changes the local topology of the network. We use N-ring structure, *i.e.* a tetrahedral group consisting of N tetrahedra sharing one common edge, instead of tetrahedron to characterize the topological change and distortion of the Delaunay network. This is owing to the fact that physically N-ring structures are presumed to be closely related to glass transition in disordered materials as 5-ring structures being considered as the glass order and other N-ring structures as disclination defects in not highly polydisperse spherical particle systems [26]. Each topological change, or flip process [22], corresponds to the creation or destruction of N-ring defects while its relationship with tetrahedron is less straightforward. N-ring analysis can therefore help establish a physically intuitive understanding of the topological change or rearrangement induced global volume variation based on creation or destruction of N-ring defects. Moreover, since each N-ring structure is simply an assembly of tetrahedra, so analysis of the distortion process of Delaunay network remains the same based on either N-ring structure or tetrahedron.

We first show how the topological rearrangement and distortion of N-ring structures leads to the global volume variation of our system. As shown in Fig. 2(a), upon shear, the fraction of different N-ring structures evolves with $\gamma$. Within one cycle, the overall trend is that the fraction of 5-ring structures decreases as the system moves away from the symmetric state and the fractions of 3-, 4- and 7-ring structures increase correspondingly, while the fraction of 6-ring structures remains nearly constant during the cycle. Nevertheless, variations are less than 2% from the symmetric state. For disordered materials, excess volumes are ordinarily related to defective structures [13]. This is consistent with the observation that increase of the totoal volume of the system $\Delta V_{global}$ (Fig. 2 (b)) is correlated with the increase of defective (3-, 4-, 6-, 7-) N-ring structures (Fig. 2(a)) and the decrease of 5-ring structures. To quantify the volume change associated with the topological rearrangement, the excess volume of each type of N-ring defective structures need to be obtained. We show subsequently that the appearance of defective structures can be viewed as the excitation process goverened by certain effective temperature, similar to the framework of shear transformation zone (STZ) theory of disordered materials [11].

Since our system has reached steady state by initial preparation, we adopt the granular thermodynamic framework originally developed by Edwards and coworkers [38]. Within this framework, the volume plays the role similar to energy in ordinary thermal system and an effective granular temperature, *i.e.* compactivity $\chi$, can be obtained based on its fluctuation. We find the use of an effective granular temperature can greatly help us understand our experimental observations.

Following the method proposed by Aste [27], we obtain compactivity $\chi$ based on the free volume $v_f$ distribution of Delaunay tetrahedron, where $v_f = v_T - v_g$, $v_T$ is the volume of each Delaunay tetrahedron and $v_g$ is the corresponding volume of particles within the Delaunay tetrahedron. If the system is in thermal equilibrium based on Edwards statistics and under the constraint of constant total volume, the probability of finding a tetrahedron possessing specific free volume $v_f$ should follow an exponential distribution, as can be derived from standard statistical mechanics [27]. As shown in Fig. 3(a), indeed the exponential behavior is found in the probability distribution function (PDF) of the tetrahedron in the large free volume region. This exponential behavior remains robust within the whole shear cycle albeit its slope varies. The slope change corresponds to the variation of the compactivity $\chi$ and its evolution within one cycle is shown in the inset of Fig. 3(a). Nevertheless, the PDFs are not exponential in the small volume regime which remains unchanged upon shear. This is due to the fact that the distributions of these small volume tetrahedra are mainly governed by the mechanical stability of the local Delaunay network topology and are less related to $\chi$ [27].

Analogous to the thermal excitation of point defects in crystals, we presume that the number of different N-ring structures are governed by $\chi$ and should follow Boltzmann distribution depending on their respective excitation volume [28], *i.e.* $N_{N-ring}/N_{5-ring}$ and $-\frac{1}{\chi}$ satisfy:

$$\frac{N_{N-ring}}{N_{5-ring}} \sim \exp\left(-\frac{E_{N-ring}}{\chi}\right), \qquad (1)$$

where $N_{N-ring}$ is the number of respective N-ring structure and $E_{N-ring}$ is its corresponding mean excitation volume from the 5-ring ground state in our system. In Fig. 3(b), we plot $N_{N-ring}/N_{5-ring}$ as functions of $-\frac{1}{\chi}$ on a semi-logarithmic plot. All relationships are clearly linear which is compatible with the existence of the Boltzmann distribution and the thermal excitation scenario. We extract the excitation volume $E_{N-ring}$ of each type of N-ring structure by fitting the slopes of the curves and obtain the corresponding excitation volume for 3-, 4-, 5-, 6- and 7-ring structures as 0.023, 0.008, 0, 0.004 and 0.023 $d^3$ respectively. After we obtain $E_{N-ring}$, we then calculate the total excitation volume resulting from the population change of different N-ring structures: $\Delta V_{topo} = \sum E_{N-ring} \times \Delta N_{N-ring}$, where $\Delta N_{N-ring}$ is the number change of each type of N-ring structure from the symmetric state. The result is shown in Fig. 2(b) and counterintuitively, we find that only 10% of totally volume change $\Delta V_{global}$ originates from $\Delta V_{topo}$.

The rest volume change instead results from the Delaunay network distortion induced mean free volume change of N-ring structures. We calculate the normalized mean N-ring structure free volume $\langle V_f \rangle_{N-ring}$ by dividing the mean free volume of each type of N-ring structure with its respective mean total volume. As shown in Fig. 3(d), we find that $\langle V_f \rangle_{N-ring}$ depend linearly on $\chi$ for all N-ring structures. Other than the intercept differences between different lines, the slopes are almost the same. If we simply fix the number of different N-ring structures during shear and only monitor the total volume change using above linear relationship with $\chi$, we find it makes up the 90% volume not accounted for by topological rearrangements. We note that this behavior is not accidental and originates from the behaviors of constituting tetrahedra. As shown in Fig. 3(c), we plot the PDFs of volume $v_f$ of tetrahedra which make up each type of N-ring structure. We find that in small $v_f$ region, the PDFs of all N-ring structures do not change with $\chi$; in large $v_f$ region, the PDFs of all N-ring structures exhibit same slope exponential tails governed by $\chi$ (inset of Fig. 3(c)). As $\chi$ increases upon shear, the fractions of tetrahedra possessing large $v_f$ increase which leads to the increase of mean free volume of N-ring structures.

Physically, the variation of mean free volume of N-ring structures comes from a contact-friction-induced fluctuation effect: upon shear, due to the presence of contact friction, the topological rearrangement among different N-ring structures does not happen instantaneously [19,22]. Instead, the N-ring structures will gradually distort before a topological rearrangement occurs. These friction-stabilized highly distorted structures can induce large volume fluctuations, as is clear from above observation that the average volume change among different N-ring structures (topological rearrangements) is actually quite small as compared to the friction-induced distortion effect. To quantify the magnitude of this fluctuation effect, we find that the dimensionless effective temperature of our system is around 0.02 which is significantly larger than the value of other sheared disordered systems [29,30] (normally on the order of $10^{-3}$). Specifically, the dimensionless effective temperature is obtained by normalizing the energy scale of the effective temperature $p\chi = 0.02 pd^3$, by the energy required to move one particle by its own diameter which is about $pd^3$, where $p$ is pressure. The one order of magnitude difference suggests that friction can significantly enhance fluctuations in granular systems and should help explain why granular materials exhibit much larger shear dilatancy effect than frictionless hard sphere system [31] and other disordered materials [29,30], since topological rearrangement is the only mechanism for dilation for those materials.

Other than volume, we also seek the microscopic structural origin of macroscopic shear force response through the evolution of the Delaunay network. For disordered system, macroscopic shear stress are normally related to the plastic processes induced by the topological rearrangements [32]. However, in sheared granular materials, energy dissipation happens both as the local structure undergoes a topological rearrangement and frictional contact evolution [33], the macroscopic mechanical response can only be reasonably understood after considering both processes. In the following, we show that an empirical constitutive relationship can well explain the global shear force variation using microscopic structural information including both topological rearrangement of the Delaunay network and contact fabric evolution induced by Delaunay network distortion.

Local topological rearrangements not only induce volume fluctuations, they also relax the stress in the system and therefore should be strongly correlated with macroscopic shear force. This correlation has been identified previously as we found that topological rearrangements or flip events mainly occur along principal stress directions [22]. Before establishing the relationship between topological rearrangement and shear force $F$, we note that both of them have orientations and a full analysis will need a tensorial description. To develop a full thermodynamic theory for the shear force, a tensorial effective temperature, *e.g.* angoricity, based on microscopic stress fluctuations [34], has to be defined, which is clearly beyond the capability of our current X-ray technique. Instead, we simplify this tensorial problem into a scalar one by using a two-state approximation analogous to previous studies [32]. Specifically, we classify all topological rearrangements into two orientation groups according to whether the angle between their orientations and one principal stress direction is smaller than with the other one. We then assume that there exists a simple proportional relation between the number of topological rearrangements with principal stress. Since shear stress is the difference between two principal stresses, we also calculate the net topological rearrangement number curve $N_{topo}$ by calculating the number difference of the two topological rearrangement orientation groups, as shown in Fig. 4(a) (see Supplemental Materials [23] for more details). Comparing $N_{topo}$ curve with the shear force $F$ evolution (Fig. 4(a) and (c)), we find that the significant hysteretic behavior of $N_{topo}$, the rapid increase of $N_{topo}$ just before shear reversal, and the abrupt drop of it after shear reversal, as shown in Fig. 4(a), are analogous to the behavior of shear force $F$ in Fig. 4(c). However, $N_{topo}$ does not match the upward sloping trend of $F$.

To account for this disparity, we also have to take Delaunay network distortion induced contact level variation into account. It is reasonable to presume that the stress variation is linearly related to the variation of contact network as normally characterized by fabric tensor [21,35,36]:

$$A = \frac{1}{N_c}\sum_{\alpha=1}^{N_c} n^\alpha \otimes n^\alpha - \frac{1}{3}I, \qquad (2)$$

where $N_c$ is the total number of contacts, $n$ is the unit contact normal vector from center to center of two particles in contact and $I$ is the unit tensor. This analysis only applies for particles with at least two contacts. We notice that the contact evolution of the system is mainly in $A_{xz}$ component. Therefore, we only use this component to approximate the contact evolution of the system, and its upward sloping trend is consistent with the behavior of $F$, as shown in Fig. 4(b).

To account for the complex shear force behavior within one cycle, we assume that both particle level topological rearrangement and contact evolution contribute linearly to the shear force $F$ to the first approximation. We adopt an empirical fitting procedure to understand their individual contributions. We normalize the $N_{topo}$ and $A_{xz}$ curves by setting the variation ranges of them to be unity and do the following fitting:

$$F = a \times N_{topo} + b \times A_{xz}. \qquad (3)$$

The fitting result is shown in Fig. 4(c) with the corresponding fitting coefficients $a = 47.4N$ and $b = 36.7N$, and their respective contributions are shown in Fig. 4(d). From the fitting, it is clear that the shear force behavior can be well accounted for after taking into account both topological rearrangement and contact evolution information. Specifically, $N_{topo}$ can explain the significant hysteresis of $F$ and its rapid variation around the reversal points, while $A_{xz}$ plays more important role in matching the upward sloping trend of $F$.

In summary, by taking friction-induced effect into consideration, combined with the knowledge of particle level topological rearrangements, we can explain the highly transient global volume and the shear force variations of a cyclically sheared granular system. These experimental findings reconcile previous theoretical approaches focusing only on particle or contact scales physics and therefore is a significant step forward to the understanding the flow of dense granular materials. It is worth noting that in our system, the majority 90% volume variation is caused by the contact-friction-induced Delaunay network distortion and the rest 10% is induced by topological rearrangements. The importance of the contact friction effect is also reflected in the contribution of contact fabric tensor to the empirical constitutive relationship. However, it is obvious that the specific fraction of contributions from these two microscopic mechanisms should be strongly correlated with the microscopic friction coefficient of granular particles which will be system dependent. It will be interesting to investigate the relative importance of two mechanisms by systematically varying the friction coefficients in future studies.

The work is supported by the National Natural Science Foundation of China (No. 11974240, No. 11675110, No. U1738120), Shanghai Science and Technology Commission (No. 19XD1402100).

* Corresponding author

yujiewang@sjtu.edu.cn

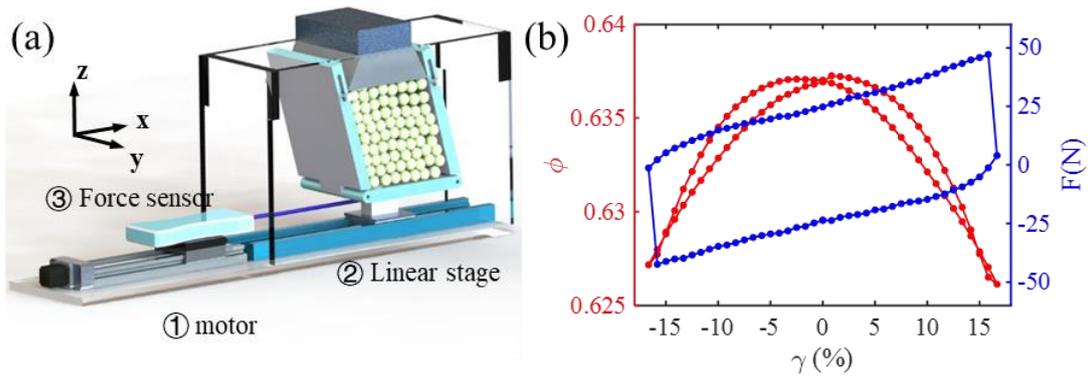

FIG. 1. (a) Schematic of the experiment setup: granular packing is sheared by the motor on the linear stage, and the macroscopic shear force is measured by the force sensor. (b) Evolutions of the average volume fraction $\phi$ and shear force $F$ within one cycle (ensemble averaged over 12 realizations).

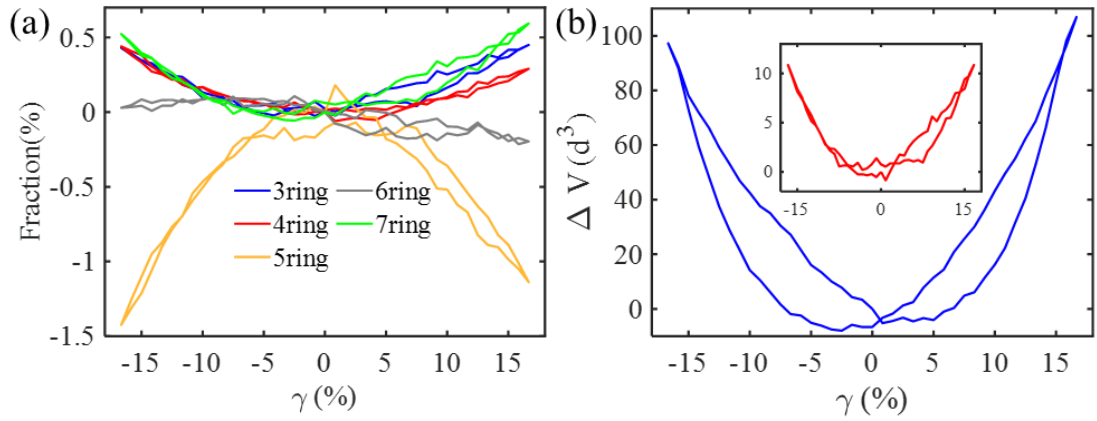

FIG. 2. (a) Variation of fractions of N-ring structures within one cycle with respect to the symmetric states ($\gamma = 0$). (b) Volume evolution of $\Delta V_{global}$ and $\Delta V_{topo}$ (inset) within one cycle.

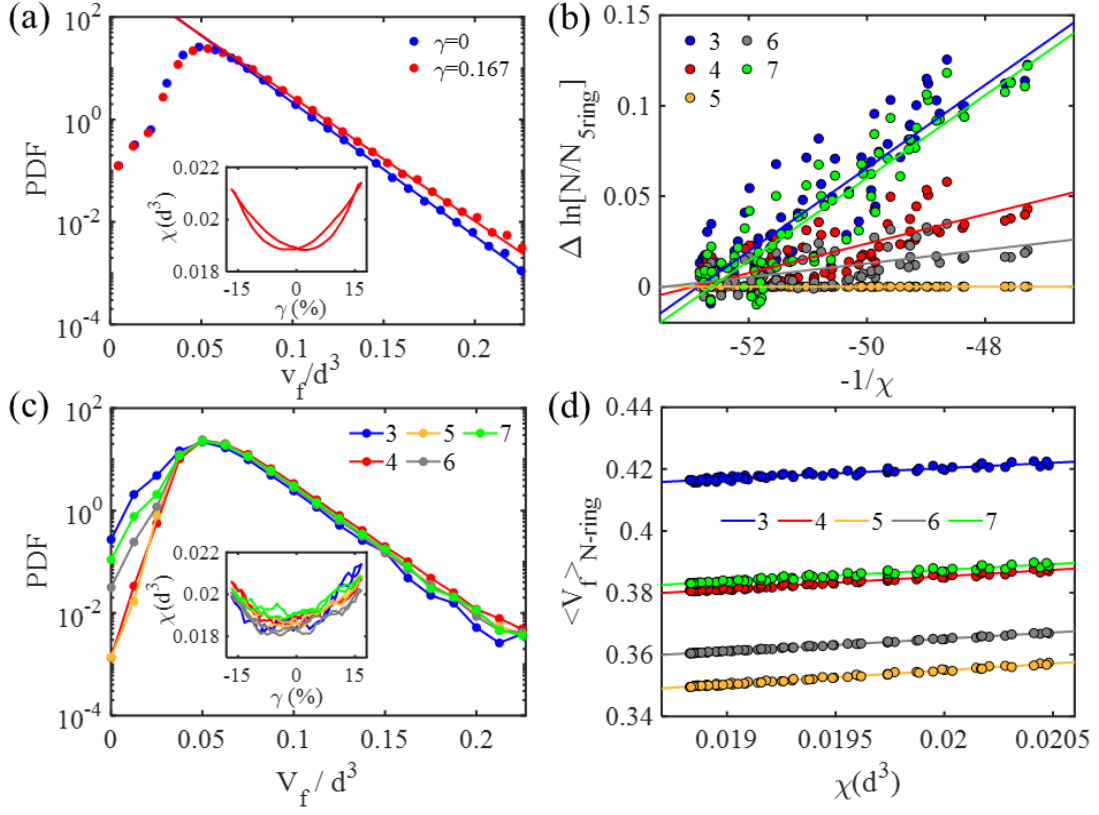

FIG. 3. (a) Probability distribution functions of the Delaunay tetrahedron free volume $v_f$ at $\gamma=0$ and $\gamma=0.167$. The solid lines denote the exponential fittings on tails. Inset: evolution of the effective temperature $\chi$ within one cycle. (b) Relationship between $\Delta ln\left(\dfrac{N_{N\text{-}ring}}{N_{5\text{-}ring}}\right)$ and $-\dfrac{1}{\chi}$. All lines are shifted vertically to have zero intercept on the vertical axis. (c) Probability distribution functions of Delaunay tetrahedron free volume $v_f$ in different N-ring structures at $\gamma=0.167$. Inset: evolution of effective temperatures for different N-ring structures within one cycle which are obtained from exponential fittings of their respective tails in the main plot. (d) Relationship between $\langle V_f \rangle_{N-ring}$ and $\chi$ for different types of N-ring structures.

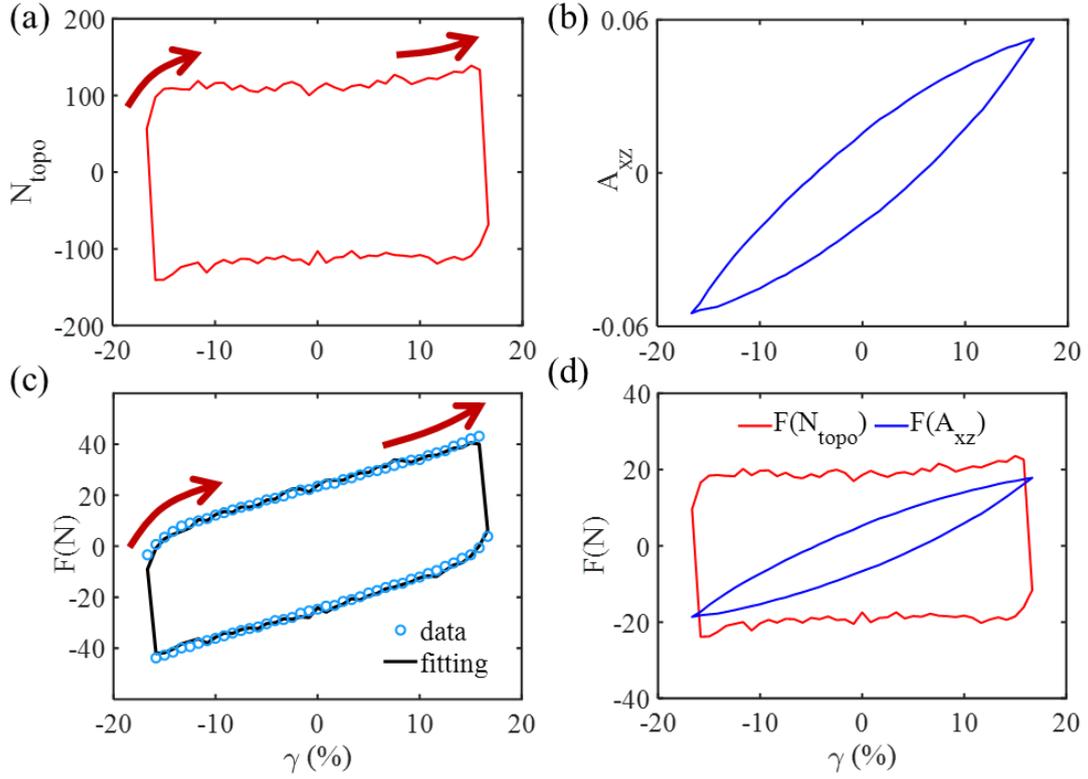

FIG. 4. (a) Evolution of $N_{topo}$ within one cycle. The arrows are guides to eye to show the increase trends of $N_{topo}$ just before and after shear reversal which is analogous to the behavior of shear force $F$ in (c). (b) Evolution of the $A_{xz}$ component of the fabric tensor within one cycle. (c) Shear force results and the corresponding fitting using $N_{topo}$ and $A_{xz}$. (d) The respective contributions of $N_{topo}$ and $A_{xz}$ to shear force $F$, denoted by $F(N_{topo})$ and $F(A_{xz})$.

# Supplemental Materials for

# X-ray tomography investigation of cyclically sheared granular materials

Yi Xing[1], Jie Zheng[1], Jindong Li[1], Yixin Cao[1], Wei Pan[1], Jie Zhang[1,2] and Yujie Wang[1,3*]

[1]*School of Physics and Astronomy, Shanghai Jiao Tong University, 800 Dong Chuan Road, Shanghai 200240, China*

[2]*Institute of Natural Sciences, Shanghai Jiao Tong University, Shanghai 200240, China*

[3]*Materials Genome Initiative Center, Shanghai Jiao Tong University, 800 Dong Chuan Road, Shanghai 200240, China*

\* Corresponding author

yujiewang@sjtu.edu.cn

**Experimental setup**

As shown in Fig. 1(a) in the main text, the shear cell is made up of five acrylic plates. The front and back plates are permanently fixed on the apparatus base. The bottom plate is mounted on a linear stage which can translate horizontally to shear the packing inside. The lower edges of the two side plates are hinged on the bottom plate while the upper edges of them are bolted on vertical slots on the front and back plates. This design allows the two side plates to adjust height freely when their lower edges are translating. The top lid is also constrained by linear guides to make sure that only the vertical motion is allowed. This guarantees that the pressure is uniformly applied on the packing when its volume varies.

The cyclic shear starts as driven by the stepping motor when the cell is in the original cubic shape, then the bottom plate is displaced in one direction by $4d$ to tilt the cell into a parallelepiped with a shear strain $\gamma = 0.167$. Subsequently, the movement is reversed until the system reaches the symmetrically tilted shape with shear strain $\gamma = -0.167$. Finally, the movement is reversed again and the cell returns back to its original cubic shape to complete one shear cycle. To make sure that the setup itself does not generate too much inner friction, we also measure the shear force of the experimental setup without sample and find it is negligible. The whole apparatus is mounted inside a CT scanner to monitor the packing's structural evolution upon shear.

**Imaging processing and in-house particle tracking algorithm**

Based on the X-ray data, we can extract the three-dimensional structural information of our granular system. To obtain the particle level structural and contact geometry information, the position and size of each particle need to be determined accurately. We follow the similar imaging processing analysis of our previous studies [22], which includes mainly the imaging binarization and maker-based watershed segmentation procedures. This imaging processing analysis ensures the accuracy of our experimental data which is estimated to be less than $3 \times 10^{-3} d$ for both the radii and centers of all the particles.

Upon shear, the structure of the system changes continuously and the particles' displacements can be obtained through a tracking algorithm based on the segmentation result of consecutive CT scans: one particle in the first scan is considered to be the same particle in the second scan which has the closest spatial location to it. Due to the congruent nature of all particles, the tracking algorithm works only if the displacement of each particle is no larger than $1/2d$ which is guaranteed by the small shear step in our experiment. In reality, the actual displacements of all particles are less than $1/3d$.

**Calculation of the net topological rearrangement number curve $N_{topo}$**

The local topological rearrangement of Delaunay network induces dissipation in the system and therefore is strongly correlated with macroscopic shear stress, which has been identified previously as we found that topological rearrangements or flip events mainly occur along principal stress directions [22]. The elementary topological rearrangement process is the neighbor switching of two coplanar tetrahedra: two neighboring coplanar tetrahedra can split into three coplanar tetrahedra, or vise versa, which is termed as 2-3 (or 3-2 flip) [22]. Consistent with our previous study [22], in our system, 2-3 and 3-2 flips mainly occur along the expansion / compression stress directions, as shown in Fig. S1.

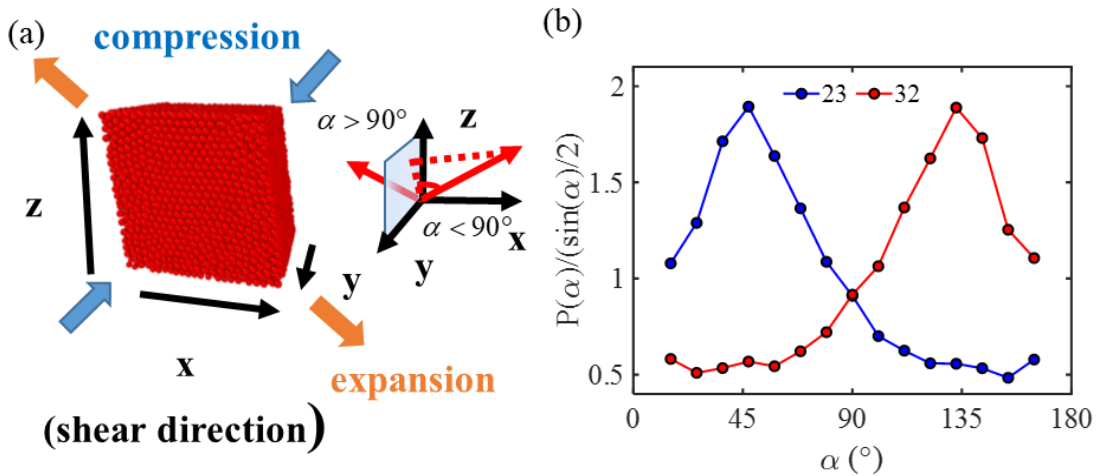

FIG. S1. (a) Schematic illustration of principle shear stress directions. The shear direction is along positive x axis, $\alpha$ is the angle between flip events and the shear direction. (b) Orientations of 2-3 and 3-2 flip events.

Before establishing the relationship between topological rearrangement and shear force $F$, we note that both of them have orientations and a full analysis will need a tensorial description. Here, we simplify the problem by using a two-state approximation similar to previous studies [32].

Following this approach, we first calculate the angles between the orientations of all flip events and two principle stresses and then classify the flip events into two types, along expansion or compression direction, according to whether the angle between each flip event and one specific principle stress direction is smaller than the other perpendicular principle stress direction. The results are shown in Fig. S2. When shear direction is reversed, the compression and expansion principle stress directions switch and all flip events change their orientations accordingly. Nevertheless, their numbers remain roughly the same as that in the opposite shear direction.

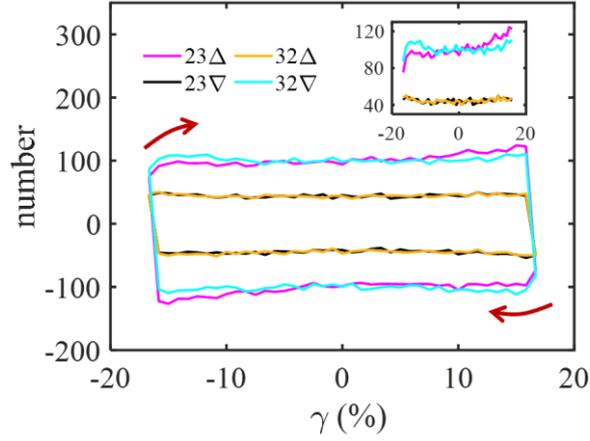

FIG. S2. The number of expansion ("$\Delta$") and compression ("$\nabla$") 2-3/3-2 events. The arrows show the directions of the loops. When shear direction is reversed, all flip events change their orientations, we use negative values of flip events to represent their orientations' changes, similar to the case of shear force. Inset: the first half cycle classification results.

Additionally, 2-3 and 3-2 flips along the same orientations are inverse processes by definition, the net flip events in either the expansion (denoted by "$\Delta$") and compression (denoted by "$\nabla$") directions is calculated as the difference between these two:

$$n(\Delta) = n(23\Delta) - n(32\Delta),$$

$$n(\nabla) = n(23\nabla) - n(32\nabla),$$

where $n(\cdot)$ represents the number of flip events and the results are shown in Fig. S3.

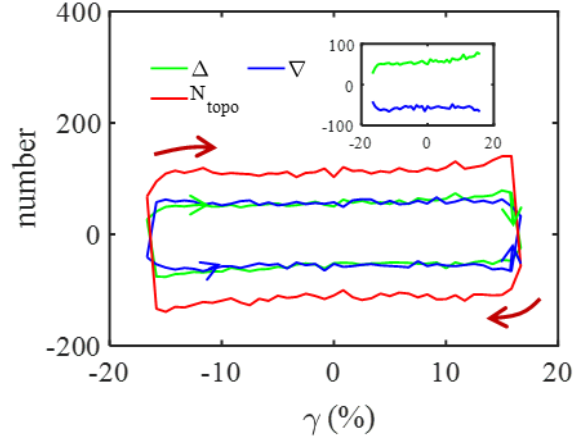

FIG. S3. The number of net expansion ("Δ"), compression ("∇") and overall topological events ( $N_{topo} = n(\Delta) - n(\nabla)$ ). The arrows show the directions of the loops. When the expansion ("Δ") curve is larger than 0, the compression ("∇") curve is negative, which leads to the opposite directions of their respective loops as shown in the main plot. Inset: the first half cycle results of net expansion ("Δ"), compression ("∇") events.

Since shear stress is proportional to the difference of the principal stresses and for simplicity, we also assume there exists a simple proportional relation between flip events and principal stress. Therefore to compare with shear stress, we calculate the net topological rearrangement number curve $N_{topo}$ by subtracting the expansion ("Δ") flip number by the compression ("∇") flip number, *i.e.*, $N_{topo} = n(\Delta) - n(\nabla)$, as shown in Fig. S3.